\documentclass[12pt]{article}

\usepackage[letterpaper,hmargin=1in,vmargin=1in]{geometry}

\usepackage{graphicx,epstopdf,amsmath,amsfonts}

\def\be{\begin{equation}}
\def\ee{\end{equation}}
\def\ba{\begin{eqnarray}}
\def\ea{\end{eqnarray}}
\def\ge{\mathrel{\raise.3ex\hbox{$>$\kern-.75em\lower1ex\hbox{$\sim$}}}}
\def\la{\mathrel{\raise.3ex\hbox{$<$\kern-.75em\lower1ex\hbox{$\sim$}}}}

\def\simgt{\mathrel{\raise.3ex\hbox{$>$\kern-.75em\lower1ex\hbox{$\sim$}}}}
\def\simlt{\mathrel{\raise.3ex\hbox{$<$\kern-.75em\lower1ex\hbox{$\sim$}}}}

\newcommand{\bi}[1]{\bibitem{#1}}
\newcommand{\fr}[2]{\frac{#1}{#2}}

\newcommand{\dsl}{\slash{\partial}}

\newcommand{\nc}{\newcommand}

\nc{\gone}{\bar g_{\pi NN}^{(1)}}
\nc{\gzero}{\bar g_{\pi NN}^{(0)}}
\nc{\al}{\alpha}
\nc{\ga}{\gamma}
\nc{\de}{\delta}
\nc{\ep}{\epsilon}
\nc{\ze}{\zeta}
\nc{\et}{\eta}
\nc{\ka}{\kappa}
\nc{\rh}{\rho}
\nc{\si}{\sigma}
\nc{\ta}{\tau}
\nc{\up}{\upsilon}
\nc{\ph}{\phi}
\nc{\ch}{\chi}
\nc{\ps}{\psi}
\nc{\om}{\omega}
\nc{\Ga}{\Gamma}
\nc{\De}{\Delta}
\nc{\La}{\Lambda}
\nc{\Si}{\Sigma}
\nc{\Up}{\Upsilon}
\nc{\Ph}{\Phi}
\nc{\Ps}{\Psi}
\nc{\Om}{\Omega}
\nc{\ptl}{\partial}
\nc{\del}{\nabla}
\nc{\ov}{\overline}
\nc{\newcaption}[1]{\centerline{\parbox{15cm}{\caption{#1}}}}

\def\beq{\begin{equation}}
\def\eeq{\end{equation}}
\def\bmat{\begin{displaymath}}
\def\emat{\end{displaymath}}
\def\bear{\begin{eqnarray}}
\def\eear{\end{eqnarray}}
\def\ba{\begin{eqnarray}}
\def\ea{\end{eqnarray}}
\def\bery{\begin{array}}
\def\ery{\end{array}}
\def\bit{\begin{itemize}}
\def\eit{\end{itemize}}
\def\ben{\begin{enumerate}}
\def\een{\end{enumerate}}
\def\btab{\begin{tabular}}
\def\etab{\end{tabular}}
\def\btbl{\begin{table}}
\def\etbl{\end{table}}
\def\bfig{\begin{figure}[htb]}
\def\efig{\end{figure}}
\def\bpic{\begin{picture}}
\def\epic{\end{picture}}

\def\ga{\mathrel{\raise.3ex\hbox{$>$\kern-.75em\lower1ex\hbox{$\sim$}}}}
\def\la{\mathrel{\raise.3ex\hbox{$<$\kern-.75em\lower1ex\hbox{$\sim$}}}}
\def\gappeq{\mathrel{\rlap {\raise.5ex\hbox{$>$}}
{\lower.5ex\hbox{$\sim$}}}}
\def\lappeq{\mathrel{\rlap{\raise.5ex\hbox{$<$}}
{\lower.5ex\hbox{$\sim$}}}}

\def\gyr{{\rm \, G\kern-0.125em yr}}
\def\mev{{\rm \, Me\kern-0.125em V}}
\def\gev{{\rm \, Ge\kern-0.125em V}}
\def\tev{{\rm \, Te\kern-0.125em V}}

\def\slash#1{\rlap{\hbox{$\mskip 1 mu /$}}#1}%

\begin{document}

\begin{titlepage}

\setcounter{page}{1}
%\begin{flushright}
%FTPI-MINN-07/35 \\
%UMN-TH-2626/07 \\
%November 2007 \\
%\end{flushright}

\vspace*{0.2in}

\begin{center}

\hspace*{-0.6cm}\parbox{17.5cm}{\Large \bf \begin{center}
Secluded WIMP Dark Matter
\end{center}}

\vspace*{0.5cm}
\normalsize

{\bf  Maxim Pospelov$^{\,(a,b)}$, Adam Ritz$^{\,(a)}$ and Mikhail Voloshin$^{\,(c,d)}$}

\smallskip
\medskip

$^{\,(a)}${\it Department of Physics and Astronomy, University of Victoria, \\
     Victoria, BC, V8P 1A1 Canada}

$^{\,(b)}${\it Perimeter Institute for Theoretical Physics, Waterloo,
ON, N2J 2W9, Canada}

$^{\,(c)}${\it William I.\ Fine Theoretical Physics Institute,\\
University of Minnesota, Minneapolis, MN~55455, USA}

$^{\,(d)}${\it Institute of Theoretical and Experimental Physics, Moscow, 117218, Russia}

\smallskip
\end{center}
\vskip0.2in

\centerline{\large\bf Abstract}

We consider a generic mechanism via which thermal relic WIMP dark matter may be
decoupled from the Standard Model, namely through  a combination of WIMP annihilation to
metastable mediators with subsequent delayed decay to Standard Model states. We illustrate
this with explicit examples of WIMPs connected to the Standard Model by metastable bosons or fermions. 
In all models, provided the 
WIMP mass is greater than that of the mediator, it can be secluded from the Standard
Model with an extremely small elastic scattering cross-section on nuclei and rate for direct 
collider production. In contrast,
indirect signatures from WIMP annihilation are consistent with a weak scale cross-section
and provide potentially observable $\gamma$-ray signals. We also point out that 
$\gamma$-ray constraints and flavor physics impose severe restrictions on MeV-scale 
variants of secluded models, and identify limited classes that pass all the 
observational  constraints.

\vfil
\leftline{November 2007}

\end{titlepage}

\section{Introduction}

The overwhelming astrophysical and cosmological evidence for dark matter has in recent years led to a dramatic
expansion in experimental programs that aim to detect observational signatures of its non-gravitational interactions
\cite{Fuller:2007hk}. These probes range from direct production at colliders to the recoil of galactic dark matter on
nuclei in underground detectors, and indirect detection of annihilation products, primarily $\gamma$-rays. A driving
paradigm in developing these probes is that of WIMP (weakly interacting massive particle) dark matter, which 
represents a simple and attractive candidate through the fact that a thermal relic with weak scale mass and annihilation
cross-section into Standard Model (SM) states naturally provides roughly the correct cosmological abundance \cite{LW}.
This weak-scale annihilation cross-section, when reversed, naturally suggests a weak-scale production cross-section
at colliders, and when viewed in the $t$-channel implies an elastic scattering cross-section on nuclei which
is within reach of purpose-built underground detectors. In recent years, experiments of the latter type have reached 
an impressive level of sensitivity \cite{CDMS,Xenon}, and significant future progress in this direction is anticipated.

Along with direct scattering on nuclei, and the neutrino signal from the annihilation of WIMPs inside the solar core, 
which is sensitive to the WIMP trapping rate (and again to the elastic scattering cross section), the
indirect detection of WIMPs via the products of their annihilation in the center of the galaxy is a distinct 
possibility \cite{Indirect}. This annihilation signal can be very model dependent, due to uncertainties in the halo profile, 
varying branching ratios, and the presence or absence of detectable monoenergetic photons. Thus, prior to specifying 
a particular model, it is impossible to say which strategy, direct or indirect detection, will provide a 
more sensitive probe of WIMPs. However, within the prevailing WIMP paradigm as exemplified by the lightest superpartner (LSP) in the MSSM
for example, the stringent constraints on the elastic scattering cross-section often impose significant limits on the 
available indirect signal from annihilation.

In this paper, we will revisit this aspect of WIMP physics, and question the commonplace assumption of a close
link between the cross-sections relevant for direct and indirect detection. More precisely, we consider what constraints
the picobarn annihilation cross-section required of a thermal relic WIMP can actually impose on its interactions with
normal matter, e.g. production at colliders or scattering off nuclei. We observe that in relatively simple models, the
latter interactions can be highly suppressed and thus in many cases the indirect annihilation signature will be the
most important probe of the non-gravitational interactions of dark matter.

\begin{figure}
\centerline{\includegraphics[width=6cm]{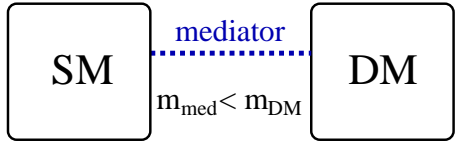}}
 \caption{\footnotesize  The secluded WIMP dark matter scenario.}
\label{f1} 
\end{figure}

Generically, any WIMP dark matter model can be conveniently decomposed into three sectors,
the SM, the WIMP itself, and the fields which mediate the WIMPs interations with the SM,
\be
{\cal L} = {\cal L}_{\rm SM} + {\cal L}_{\rm WIMP} + {\cal L}_{ \rm mediator},
\label{L}
\ee
and in many models the mediator states are in fact part of the SM, e.g. the electroweak 
gauge bosons or the Higgs. In this paper, we point out a simple and generic mechanism that 
allows the WIMP, while still a thermal relic,  to be {\it secluded} from the SM, dramatically reducing its couplings to SM states, and 
consequently suppressing the collider and direct detection rates by many orders of magnitude. 
The mechanism relies on a metastable mediator that couples the SM to the secluded WIMP sector, with a mass
less than that of the WIMP, see Fig.~\ref{f1}. In this kinematic regime,  direct annihilation into 
a pair of mediators is always possible. Due to their coupling to the Standard Model, these particles
are unstable but the constraint on their lifetime is very weak, and in particular a lifetime under
one second is sufficient to guarantee their decay before the beginning of primordial nucleosynthesis (BBN),
rendering them completely harmless. Secluded WIMP models may therefore be 
impossible to detect using colliders or direct searches, but the indirect signatures, through 
e.g. annihilation in the Galactic center, can be as pronounced as in any WIMP scenario.

To illustrate this mechanism, in Section~2 we will construct several models with fermionic, 
scalar and vector particles as mediators. We show that if $m_{\rm WIMP}<m_{\rm mediator}$, the parameter space of
such models is highly constrained,
as the coupling of the mediator to the SM  must necessarily be sizable to
ensure the required annihilation cross-section. Yet if the reverse is true, $m_{\rm WIMP}>m_{\rm mediator}$, there are no strict requirements on the 
size of the mixing except for the lifetime of the mediator state, which in some instances can be satisfied for
(mixing)$^2$ of the mediator with the SM as low as $10^{-23}$. 

An interesting limiting regime of the secluded scenario arises when the mediator (and the WIMP) are both very light.
This ties in with models of MeV-scale dark matter 
that provide a tantalizing yet speculative link \cite{511} to the unexpectedly strong 511 keV line observed from the galactic center \cite{Integral}. 
We show in Section~3 that some secluded WIMPs may have advantages over existing models of this type, 
and discuss their observational signatures at low energies, including 
missing energy signals in meson decays.

\section{Models of secluded WIMPs}

For some time the discussion of WIMPs has centered around supersymmetric models where the 
LSP is stable, provided that $R$-parity is unbroken. Supersymmetry is largely motivated 
by a well-known combination of theoretical arguments unrelated to dark matter, and the possible existence of a 
stable or long-lived WIMP-LSP  may provide an interesting bridge to cosmology. On the other 
hand, to date there are no experimental indications of supersymmetry, while there is ample evidence for
the existence of dark matter. Therefore, an alternative approach to the particle physics of dark matter that is certainly
logical and justifiable, consists of studying generic classes of WIMP models, among which the minimal 
choices are obviously well motivated. Over the years, WIMP models with Higgs and/or singlet mediation 
have been studied extensively  \cite{singlet1,singlet2,singlet3}. More recently, models with exotic electroweak 
matter were also considered in some detail \cite{EWDM}, as well as models with additional 
gauge groups \cite{ZprimeDM}. More generally, going beyond the WIMP framework also allows for freedom
in tuning the coupling to the SM, as in scenarios with sterile neutrinos \cite{sterile} or dark matter populated by
late decays \cite{swimp}. In this paper we adhere to a rather minimalist WIMP framework, 
which is well suited to demonstrating our main point.

\subsection{U(1)$'$ mediator}

We can construct a simple secluded model, starting from the Lagrangian for a Dirac WIMP, whose 
interaction with the SM is mediated by an additional U(1)$'$ gauge group: 
\be
{\cal L}_{ \rm WIMP+mediator} = -\fr{1}{4}V_{\mu\nu}^2 - \fr{\kappa}{2}V_{\mu\nu}B_{\mu\nu} 
-|D_\mu \phi|^2 - U(\phi\phi^*) + \bar \psi (iD_\mu\gamma_\mu - m_\psi) \psi.
\label{V'}
\ee
In this Lagrangian, $\psi$ and $V_\mu$ denote respectively the Dirac WIMP and the U(1)$'$ vector boson
mediator, with field strength $V_{\mu\nu}$ and covariant derivative $D_\mu = \partial_\mu + i e' V_\mu$. 
To avoid confusion, we denote the strength of the U(1)$'$ coupling constant
as $e'$. The U(1)$'$ vector bosons $V_\mu$ couple to the SM hypercharge gauge bosons $B_\mu$ via 
kinetic mixing. The additional scalar $\phi$ higgses U(1)$'$ at or near the weak scale. Note that the SM and WIMP
sectors are coupled only via the mediator at the renormalizable level, the SM is neutral under U(1)$'$, and the WIMP sector is a singlet
under the SM gauge group. Assuming the scalar potential $U$ has a minimum away from zero, 
the Lagrangian below the U(1)$'$ breaking scale takes the form, 
\be
{\cal L}_{ \rm WIMP+mediator} = -\fr{1}{4}V_{\mu\nu}^2 +\fr{1}{2} m_V^2 V_\mu^2 + \kappa V_\nu \partial_\mu B_{\mu\nu}
+\bar \psi (iD_\mu\gamma_\mu - m_\psi) \psi +{\cal L}_{h'},
\label{VdB}
\ee
where $m_V$ is the resulting mass of the U(1)$'$ vector boson, and ${\cal L}_{h'}$ is the 
Lagrangian for the Higgs$'$ particles including their self-interaction and interactions with $V_\mu$.
 
\begin{figure}
\centerline{\includegraphics[bb=0 460 570 740, clip=true, width=10cm]{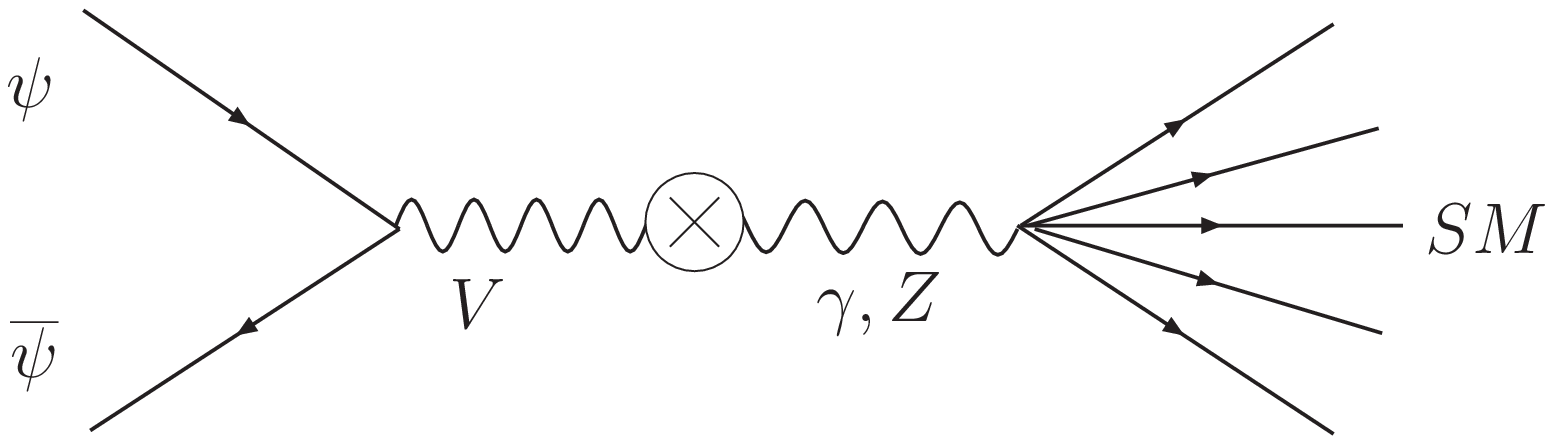}
  \includegraphics[bb=100 400 550 740, clip=true, width=7cm]{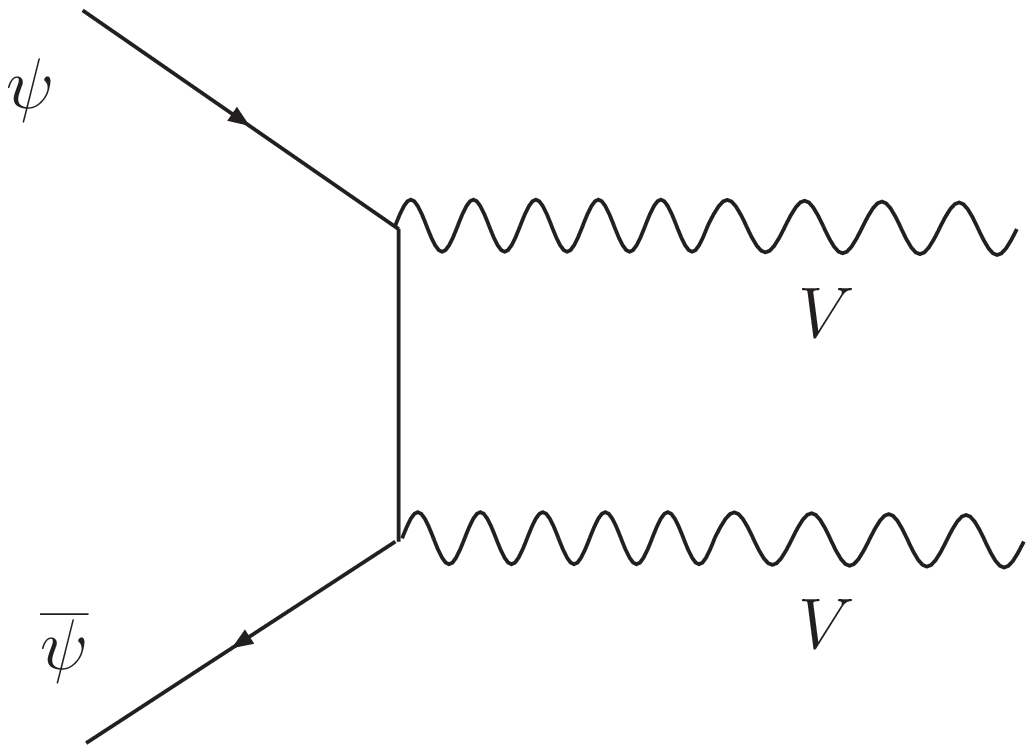}}
\vspace*{-0.6in}
 \caption{\footnotesize  WIMP annihilation for: (A) $m_{\psi} < m_V$ on the left; and (B) $m_{\psi} > m_V$ on the right -- the secluded regime in which
 the annihilation may proceed via two metastable on-shell $V$'s, which ultimately decay to SM states.}
\label{f2} 
\end{figure}

 There are four important parameters in the model,  the WIMP and vector boson masses
 $m_\psi$ and $m_V$, the mixing parameter $\kappa$ and new coupling constant $e'$. 
%All relevant observable quantities can be easily calculated in terms of these parameters.
The most important quantity for WIMP physics is arguably the annihilation cross section 
into the SM states. To this end it is easy to identify the two primary 
mechanisms responsible for annihilation (see Fig.~\ref{f2}):
\begin{enumerate}
\item[(A)] $\psi$ + $\bar\psi$ $\to$ virtual $V$ $\to$ virtual $\gamma,Z$ $\to$ SM states.
\item[(B)] $\psi$ + $\bar\psi$ $\to$ on-shell $V+V$, with subsequent decay to SM states.
\end{enumerate}

Process (B) is open only in the kinematic regime  $m_\psi > m_V$ while process (A) can occur regardless 
of the relation between $m_V$ and $m_\psi$. We will analyze the case of $m_\psi < m_V$ first.

\subsubsection{The characteristic WIMP regime}

The annihilation cross section in this case is 
given by the diagram in Figure \ref{f2}(A). Although we use the full result for numerical analysis, to simplify the presentation its helpful to quote the annihilation 
cross section in the limit $ m_Z^2, m_t^2, m_h^2   \ll m_\psi^2$. Since $2m_\psi$ then provides the energy scale for the problem, in
 this limit one may substitute $\partial_\mu B_{\mu\nu}$ by the total hypercharge current and neglect the influence of SM threshold effects.
For small mixing, characterized by $\beta \ll 1$ where 
\be
 \beta \equiv \left(\fr{\kappa e'}{e\cos\theta_W}\right)^2,
 \ee
the resulting annihilation cross section for nonrelativistic WIMPs takes the following form,
\be
\langle \sigma_{\rm ann} v\rangle_{m_\psi \gg m_{\rm SM}} \approx 1.3 ~{\rm pbn} \times \beta\left(\fr{500 ~ \rm GeV}{m_\psi}\right)^2
 \times \left(\fr{4m_\psi^2}{4m_\psi^2-m_V^2}\right)^2,
\ee
proceeding in the $l=0$ channel with an obvious pole at $m_\psi=m_V/2$, in the vicinity of 
which a more accurate treatment of the thermal average is required. The result depends on the mixing parameter
$\beta$ and the sum of squares of the hypercharges for the SM fields, 
$\sum_{\rm fermions}Y_f^2 + \fr12\sum_{\rm bosons}Y_b^2 = 10 + 0.25$.
Note that in the opposite limit, $m_b\ll m_\psi\ll m_Z$, the total cross section is instead proportional to the 
sum of squares of all the electric charges of SM fermions with the exception of the $t$-quark. 

This cross-section needs to be compared with the constraint on the dark matter energy density provided by recent 
cosmological observations:
\be
2 \times \fr{10^9(m_\psi/T_f)}{\sqrt{g_*(T_f)}\times {\rm GeV} \times M_{\rm Pl} 
\langle \sigma v \rangle } \leq \Omega_{DM} h^2 \simeq 0.1,
\label{abundance}
\ee
where $T_f$ is the freeze-out temperature (it suffices here to take $m_\psi/T_f \simeq 20$), $g_*$ the effective number of degrees of freedom at 
freeze-out, %$M_{\rm Pl} = 1.2\times 10^{19}$ GeV, 
and the extra factor of two relative to the standard formula
(see {\em e.g.} \cite{KT}) is because annihilation can occur only between particles and anti-particles.

\begin{figure}
\centerline{\includegraphics[width=7cm]{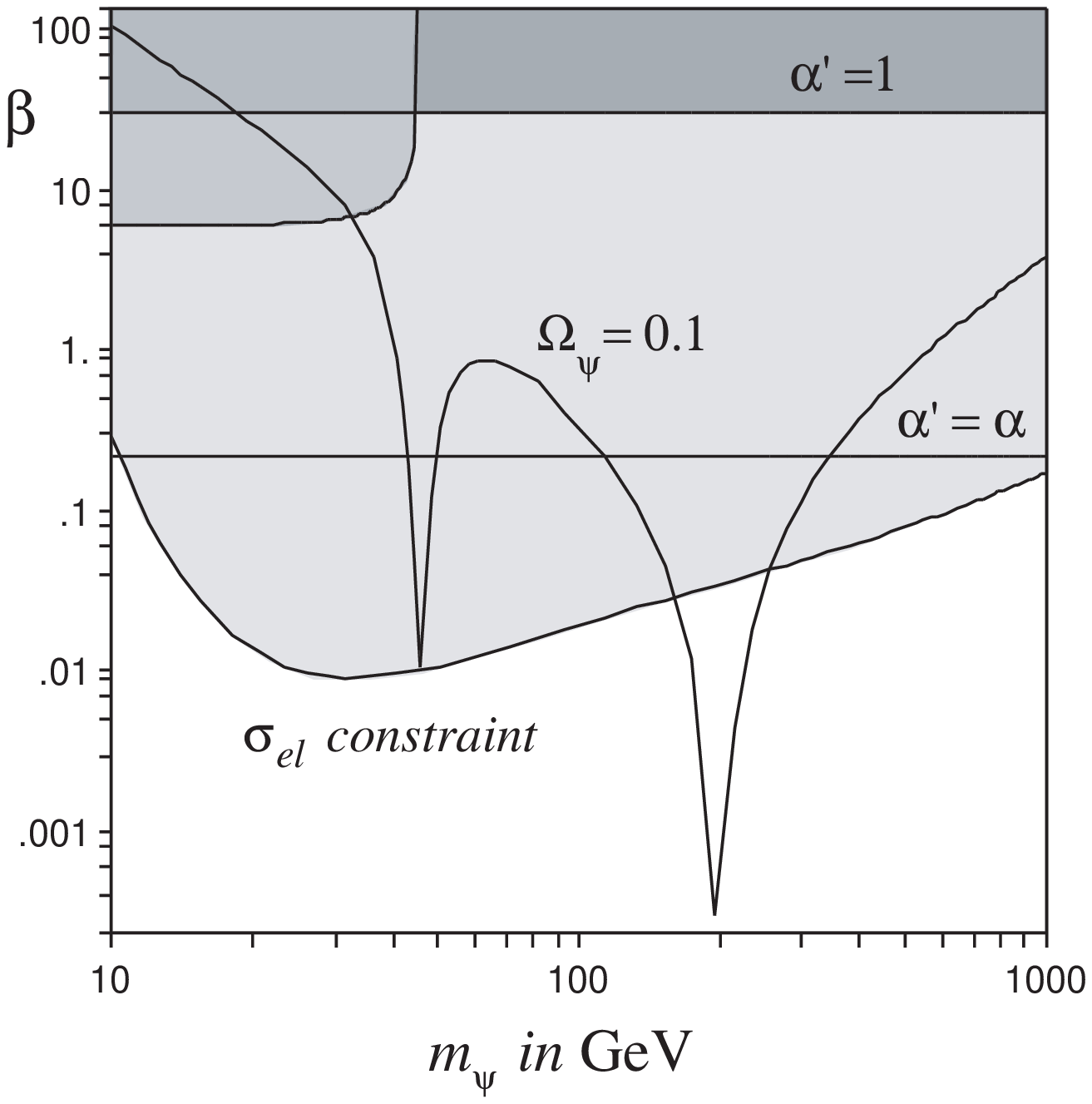}\hspace*{0.3in} \includegraphics[bb=0 -12 379 389, clip=true,width=6.95cm]{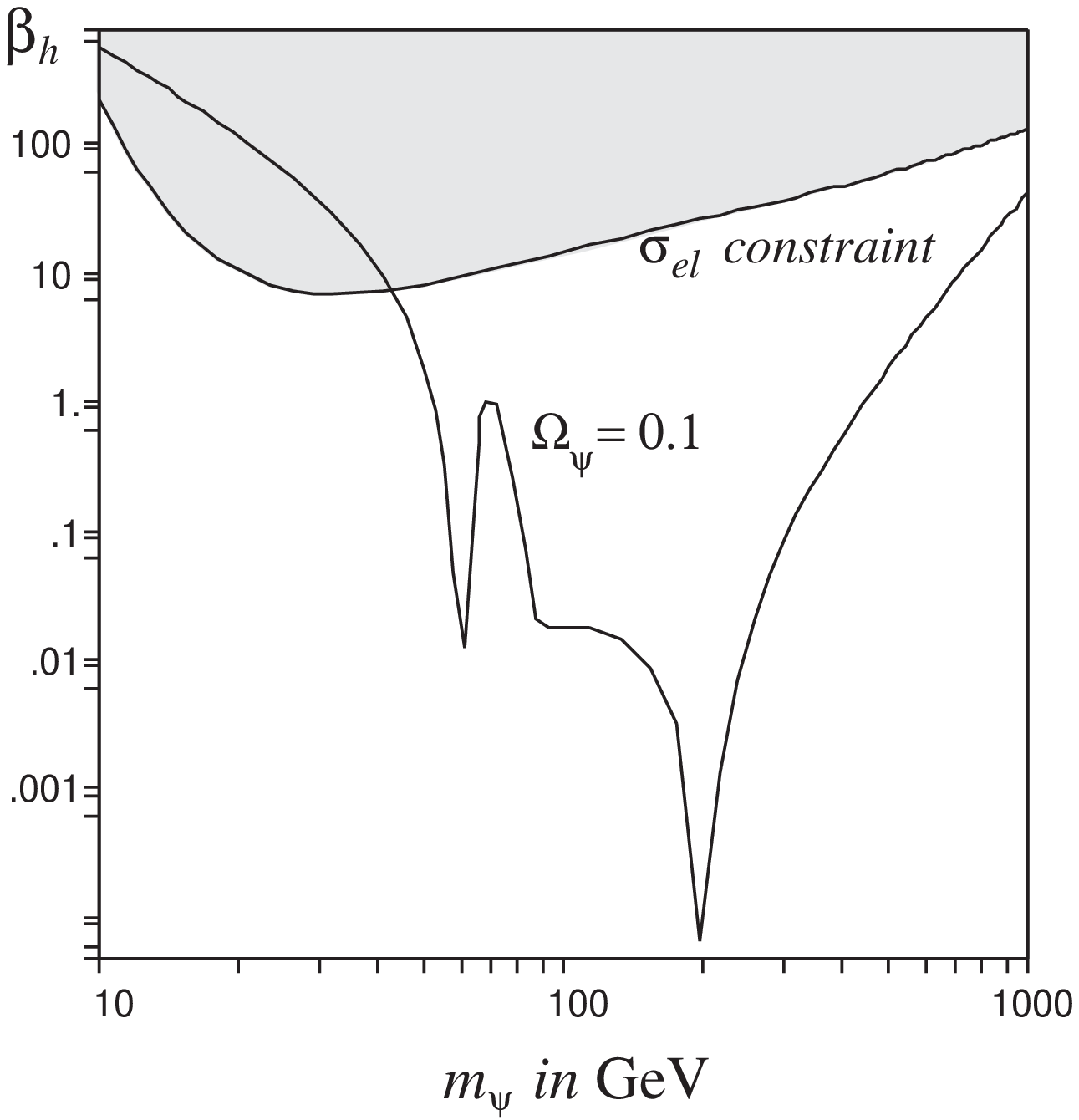}}
 \caption{\footnotesize (A) On the left, the parameter space of the $U(1)'$-mediated model. 
The elastic cross section constraint excludes 
a large portion of the $m_\psi-\beta$ parameter space. The upper-left corner is also excluded due to 
the invisible decay of $Z$ into WIMPs. The two horizontal lines correspond to the collider limits 
on four-fermion interactions with $\alpha'=1$ and $\alpha'=\alpha$. 
(B) On the right, the parameter space of the singlet-Higgs mediated model. There are no collider constraints, and the 
$\sigma_{el}$ constraint is much weaker.
}
\label{f3} 
\end{figure}

In Fig.~\ref{f3}, we exhibit the abundance constraint on the $\beta-m_\psi$ plane for a specific choice
of mediator mass, $m_V =400$~GeV, by saturating the inequality (\ref{abundance}).  This value of $m_V$ lies outside 
the direct reach of LEP or the Tevatron but is certainly within range for the LHC. One can clearly see
the enhancement of the annihilation cross section in the vicinity of the two vector resonance poles, $Z$ and $V$, 
where the mixing parameter $\beta$ is allowed to be significantly smaller than 1.

This model is subject to various constraints from direct searches and collider physics.
\begin{itemize}
\item[(a)] 
The elastic scattering of galactic WIMPs off nuclei occurs with 
a characteristic momentum transfer of order 100 MeV or less, making virtual photon exchange the dominant 
contribution to the scattering amplitude. From an appropriate low-energy effective theory viewpoint, the WIMP
$\psi$ is electrically neutral but exhibits a non-zero charge radius given by
\be
r_c^2 = 6\fr{\kappa e'}{e}\times \fr{1}{m_V^2} = 6~\fr{\beta^{1/2}\cos\theta_W}{m_V^2}.
\ee
The contribution of $r_c$ to the elastic scattering of WIMPs off nuclei was calculated 
in \cite{PtV}, and can easily be rescaled to the equivalent cross section per nucleon,
\be
\sigma_{\rm el} = \fr{4\pi}{9}Z^2\alpha^2r_c^4\left(\fr{m_Am_D}{m_A+m_D}\right)^2 \quad \Longrightarrow \quad 
\sigma_{\rm nucleon} = \fr{4\pi}{9}\alpha^2r_c^4m_p^2\left(\fr{Z}{A}\right)^2,
\ee
where $m_A$ and $m_p$ are nuclear and nucleon masses, and the $Z/A$ ratio should be specified for the 
relevant experimental setup. We plot the corresponding experimental limit recently obtained by the 
XENON collaboration \cite{Xenon} in Fig.~\ref{f3}, which clearly rules out a significant
portion of the parameter space away from the $m_\psi = M_V/2$
resonance. This is not surprising, as the model is in many ways similar to the original ``heavy Dirac neutrino"
of Ref.~\cite{LW}, which is known to be essentially excluded by direct searches.

\item[(b)] Other particle physics constraints on this model are also highly dependent on $m_V$. 
For $m_V\gg m_Z$ and $E$, where $E$ is the energy accessible in the collision, one can approximate
$V$-exchange between SM particles as an effective current-current interaction, 
\be
{\cal L}_{\rm eff} = \fr{4\pi\alpha\kappa^2 }{m_V^2\cos\theta_W^2} J^Y_\mu J^Y_\mu.
\ee
We can then constrain the coefficient using limits on the 
corresponding effective four-lepton operator from
LEP2 and Tevatron searches for ``compositeness" \cite{LEP,Tevatron}. 
In terms of the conventionally normalized coefficient,  $4\pi/\Lambda_c^2$, 
no deviation is observed from the SM cross section up to $\Lambda_c\sim$10-15 TeV. Consequently,  
we arrive at the following collider constraint,
\be
\Lambda_c < 10~{\rm TeV} \quad \Longrightarrow \quad \beta < 0.3\times\fr{\alpha'}{\alpha}\times
\left(\fr{m_V}{500~{\rm GeV}}\right)^2.
\ee 
In order to plot this constraint in Fig.~\ref{f3}, we choose two representative values for the 
coupling constant $e'$, one defining the perturbative regime, $\alpha' <1$, and a more realistic 
line for $\alpha'=\alpha$. 

\item[(c)] For $m_\psi< m_Z/2$, there is also an extra contribution 
to the invisible width of $Z$, namely $Z\to \psi\bar\psi$,
\be
\Gamma_{Z\to \psi \bar\psi} = \fr{\alpha\beta}{3} \fr{m_Z^4}{(m_V^2-m_Z^2)^2}\left(1+\fr{2m_\psi^2}{m_Z^2}\right)
\sqrt{m_Z^2-4m_\psi^2}.
\ee
Requiring this width to be less than 4 MeV \cite{PDG} results in an additional constraint plotted in 
Fig.~\ref{f3}. 
\end{itemize}

As is evident from the figure, the existing constraints already rule out WIMP masses up to 100 GeV with the
exception of a small region in the vicinity of $m_\psi = m_V/2$ where there is resonant enhancement in the 
annihilation cross section. Future LHC  experiments, and the next generation of dark matter experiments, will provide much deeper
probes into the parameter space of this model.

\subsubsection{The secluded WIMP regime}

The situation changes drastically if annihilation process (B) is kinematically allowed. 
The cross section for WIMP annihilation into pairs of (unstable) $V$ bosons is then given 
by
\be
\sigma v = \fr{\pi (\alpha')^2}{m_\psi^2}\sqrt{1-\fr{m_V^2}{m_\psi^2}},
\ee
which together with (\ref{abundance}) implies that in the limit $\beta \ll 1$
the correct dark matter abundance is achieved if
\be
\alpha'\times \left(1-\fr{m_V^2}{m_\psi^2}\right)^{1/4} \simeq 5\times 10^{-3} \times \left(\fr{m_\psi}{ {\rm 500~ GeV}}\right).
\ee
This constraint is easily satisfied for a rather natural range for $m_\psi$ and $\alpha'$. 
Crucially, since large mixing is no longer required in this kinematic regime to ensure the correct annihilation
cross-section, $\ka$ can be taken very small indeed. The only constraints one has to impose are 
that the decay of $V$ (and also $h'$)
occur before the start of nucleosynthesis. By choosing $m_h' > m_V/2$, only the decays of $V$ are 
sensitive to the mixing:
\be
\Gamma_V \geq {\rm s}^{-1} \quad \Longrightarrow \quad    \kappa^2 \left(\fr{m_V}{10~ {\rm GeV}}\right) \ga 10^{-23}.
\label{BBNkappa}
\ee
A tighter constraint would follow from requiring $V$ decays to remain in thermal equilibrium,
which would also ensure the initial thermal and chemical equilibirium for WIMPs,  
used in  the derivation of the abundance formula (\ref{abundance}), 
\be
\Gamma_V \geq {\rm Hubble~ Rate}\, [T\simeq 0.05 m_\psi] \quad \Longrightarrow \quad    \kappa^2 \left(\fr{m_V}{10~ {\rm GeV}}\right) \ga 10^{-12} 
\left(\fr{m_\psi}{500~{\rm GeV}}\right)^2.
\label{TEkappa}
\ee
Although considerably tighter than (\ref{BBNkappa}), the constraint (\ref{TEkappa}) does not change the main 
conclusion: in the limit $m_{\rm mediator} <m_{\rm WIMP}$, the 
WIMP sector can be secluded and neither collider nor underground searches 
impose any significant restrictions on the model. Moreover, the constraint (\ref{TEkappa}) can be relaxed to 
(\ref{BBNkappa}) if some new UV physics ensures proper thermal contact between the dark matter and 
SM sectors at higher temperatures resulting in $T_{DM}\sim O(0.1)\times T_{SM}$ 
at the time of dark matter annihilation. For example, a dimension-8  operator of the form 
$V_{\mu\nu}^2(F^{SM}_{\mu\nu})^2$ generated at a rather high energy scale would 
suit this purpose. 

It is also worth noting that one could completely sever the remaining link to the SM, and set the
mixing parameter $\kappa$ to zero, {\em if} the U(1)$'$ gauge symmetry is not broken at all or broken only at 
very low energy scales. In this case, there will be an extra component to the energy density of the universe, namely the
``dark radiation'' associated with massless $V$-bosons. However, this energy density associated with $V$ may be 
much smaller than that of the SM photons, because the decoupling of the dark sector may have occurred at a 
very early epoch, after which the photon temperature was effectively increased several-fold by input from the decay of numerous
SM degrees of freedom. Although at first sight this model looks like a completely decoupled dark sector, the energy density 
in $V$ could in principle be detected by highly sensitive next-generation CMB anisotropy probes at 
very small angular scales.

\subsection{Singlet scalar mediator}

An alternative class of models uses scalar mediators, discussed for example in several
recent publications \cite{singlet3}. The simplest example consists of two additional fields, the singlet 
WIMP fermion $\psi$ and a singlet mediator $\phi$,
\begin{eqnarray}
{\cal L}_{ \rm WIMP+mediator} & = &
\fr12 (\partial_\mu \phi)^2 - \fr12 m_\phi^2\phi^2 + \bar \psi (i\partial_\mu\gamma_\mu - m_\psi-\lambda_\psi \phi) \psi
\\\nonumber &-&\lambda_1 v \phi(H^\dagger H - \fr{v^2}{2}) - \lambda_2 \phi^2(H^\dagger H - 
\fr{v^2}{2}) - V(\ph).
\label{Higgsmodel}
\end{eqnarray}
Here $H$ is the SM Higgs doublet and $v$ the corresponding vacuuum expectation value, introduced in (\ref{Higgsmodel})
for convenience. By a redefinition of the $\phi$ field, we can always set
$\langle \phi \rangle =0$. Should one pursue the minimalist approach, $\phi$ itself can be a 
dark matter candidate, which would then fix the value of $\lambda_2$. However, to 
demonstrate our point as simply as possible, we take the limit $\lambda_2\to 0$. After  
electroweak symmetry breaking, the relevant dark matter Lagrangian then takes the following form,
\be
{\cal L}_{ \rm WIMP+mediator} =
- \fr12 (\partial_\mu \phi)^2 + \bar \psi (i\partial_\mu\gamma_\mu - m_\psi-\lambda_\psi \phi) \psi
-\lambda_1 v^2 \phi h +\cdots,
\label{phi-h}
\ee
where all interactions between SM fields and WIMPs are mediated by  Higgs-singlet mixing. 

The analysis of this model follows similar lines to the example above, so we will be rather brief.
We first consider the case $m_\phi> m_\psi$ and choose $m_\phi = 400 $ GeV to enable a clear comparison with the 
previous model. 
Working to lowest order in the mixing parameter, we have computed the annihilation cross section taking into account
 $t\bar t$, $ZZ$, $WW$, $hh$ and $b\bar b$ in the final state, following {\em e.g.} Ref.~\cite{singlet2}. The cross section 
 has a leading $p$-wave contribution, 
which provides an additional suppression to $\langle \sigma v \rangle.$\footnote{The presence of a 
pseudoscalar coupling, $\phi \bar\psi i\gamma_5 \psi$, would open up the $s$-channel.}
Choosing the Higgs mass to be $m_h = 120$ GeV, and denoting the mixing parameter by $\beta_h$,
\be
\beta_h \equiv \fr{\lambda^2_\psi\lambda^2_1v^4}{m_\phi^4},
\ee
we can use the abundance constraint (\ref{abundance}) to plot the dependence of 
$\beta_h$ on the WIMP mass in Fig.~3(B). This plot also includes 
the experimental constraint on the nucleon-WIMP elastic scattering cross section induced by $h-\phi$ 
scalar exchange.  We take the SM value of the Higgs-nucleon coupling to be  
 $g_{hNN} \simeq 300$ MeV. There are no significant collider constraints on this WIMP model, as all WIMP production 
 has to occur via a real or virtual Higgs or Higgs-like bosons, and this is highly suppressed. 
 Although quite constraining for low WIMP masses, in this scenario 
 the direct search experiments cannot probe $m_\psi$ above 50 GeV. 
 
 Again, the situation changes significantly if $m_\phi< m_\psi$, which allows for the WIMP sector to be secluded. 
 The kinematically allowed pair-annihilation of WIMPs into two $\phi$-scalars
 then simply imposes a constraint on a combination of $\lambda_\psi$, $m_\phi$ and $m_\psi$, but 
removes any constraints on $\lambda_1$. The decay of these scalars before BBN imposes a very relaxed requirement of
 $\lambda_1 \ga 10^{-21}$. In practice, even taking $\beta_h \sim 10^{-5}$ would eliminate any chances for direct search experiments 
 and/or colliders to probe the WIMP sector in such a model.

 \subsection{Right-handed neutrino mediator}

In previous subsections the choice of metastable mediator, although relatively simple, 
was nonetheless exotic. Neither metastable vector or scalar particles are required by any 
known SM physics. There is, however, the distinct possibility of promoting right-handed 
neutrinos $N_R$, arguably the best motivated extension of the SM field content,  
to the role of metastable mediators with a dark matter sector. Indeed, if the right-handed neutrinos 
have Majorana masses of order the electroweak scale, their decay width into SM states such as 
left-handed neutrinos and the Higgs will be proportional to the square of the Yukawa 
coupling, {\em i.e.} in practice to the light neutrino masses. As is clear from the 
discussion of the previous subsections, $\Gamma_{N_R}\sim m_\nu\sim\, 0.1\,$eV  is in the right range
for $N_R$ to play the role of a metastable mediator. To complete this model, we choose the 
secluded sector to have one additional fermion $N'$ and a boson $S$:
\begin{eqnarray}
{\cal L}_{ \rm WIMP+mediator} & = &
\fr12 (\partial_\mu S)^2 - \fr{m_S^2}{2}S^2 + \bar N'i\dsl N' - \fr{m_{N'}}{2} N'^T N'+
\bar N_Ri\dsl N_R - \fr{m_{N_R}}{2} N_R^T N_R\nonumber
\\ &-& \fr{\lambda}{2}S^2H^\dagger H -\left[Y_\nu \bar L H N_R - Y_{N'} S N_R^T N' +(h.c.)\right].
\label{RHmediator}
\end{eqnarray}
This Lagrangian possesses a $Z_2$ symmetry which allows only even numbers of $N'$ or $S$ 
at each vertex. Alternatively, one may complexify $S$ and $N'$ and introduce a new global 
charge that would ensure the same property. 

A viable secluded dark matter model results from the choice $m_S,\,m_{N'}>m_{N_R}$. 
If $m_S>m_{N'}$, then $S$-mediated annihilation $N'+N'\to N_R+N_R$ will ensure the right WIMP abundance of 
$N'$ given an appropriate choice of  $Y_{N'}$. Notice that a scalar coupling $\lambda$ as small as $10^{-8}$ is
sufficient to keep the dark sector in thermal/chemical equilibrium prior to freezeout \cite{singlet2}. For $m_S<m_{N'}$
the roles are reversed, and $N'$-mediated annihilation allows for $S$ to be dark matter. In either case, the
dark matter candidate, either $N'$ or $S$, is secluded from direct observational probes.

\subsection{Strong non-Abelian interactions in the secluded sector}

Models with a non-Abelian gauge group $G'$ in a hidden sector can also be secluded and may be of interest in
the context of dark matter models with strong self-interactions \cite{sidm} if the scale is relatively light. The gauge bosons
of $G'$ cannot couple directly to the SM  because of gauge invariance. 
Therefore, in this case the mediators must be charged under both gauge groups. The simplest 
example of this kind is given by
\be
{\cal L}_{ \rm WIMP+mediator} = -\fr{1}{4}(G^a_{\mu\nu})^2 + 
\sum_f \bar f (i\gamma_\mu D_\mu^{\rm SM,hid} -m_f)f + \sum_\psi\bar\psi(i\gamma_\mu D_\mu^{\rm hid} -m_\psi)\psi,
\label{composite}
\ee
where $G^a_{\mu\nu}$ is the non-Abelian field strength in the hidden sector, while $\psi$ and $f$ are fermions, 
with $\psi$ charged only under $G'$, while the fermions $f$ play the role of mediators and are charged under
both $G'$ and the SM gauge group. This field content will necessarily have to satisfy anomaly cancelation constraints.
 We further assume that the confinement scale of $G'$ is  comparable to or larger than the electroweak scale, $\Lambda' > v$. 

The Lagrangian (\ref{composite}) allows the construction of  a secluded WIMP model, in which both mediators and dark matter 
are composite. If all masses in the $f$-sector are large, these fields can be integrated out leading to non-renormalizable 
interactions between the two sectors of the form $\frac{1}{m_f^4}(G^a_{\mu\nu})^2 (F^{\rm SM}_{\mu\nu})^2$.
The phenomenology of such terms, in connection with metastable dark matter composed of hidden-sector glueballs was considered in Ref. \cite{FP}. 
The $\psi$-containing baryons $B'$ and antibaryons are viable dark matter candidates if the nonrenormalizable terms leading to their 
decays are forbidden. The mass of an exotic ``meson'' ($M' \sim \bar\psi \psi$) would be smaller than the baryon mass, and 
the annihilation process $B'\bar B' \to M'\bar M'$ therefore open. Since this cross section is not expected to have additional 
suppression other than that provided by the mass scale of the exotic baryons, the abundance constraint would require 
that the dark matter mass be above 10 TeV.  The mesons $M'$ as well as 
the glueballs of the hidden group would be unstable with respect to decay into the 
SM fields, with total widths controlled by the combination $\Lambda'^9/m_f^8$ \cite{FP}. 
For large $m_f$, i.e. $m_f\gg\Lambda'$,  this width can be exceptionally small, {\em e.g.} of order the 
Hubble scale during $B \bar B$ freeze-out. Explicit models with exotic baryons as dark matter
have previously been constructed within the framework of 
gauge-mediated supersymmetry breaking \cite{Dimopoulos,Hamaguchi}.

The opposite limit, $m_f \ll \Lambda'$, would provide a regime where the WIMP sector could in principle be probed 
because the mesons built from $f$ would then be of electroweak scale, and will interact both with the SM and
with WIMPs in the form of exotic baryons. Although the direct detection signal
signal might again be rather low, the exotic $f$-containing mesons could 
conceivably be produced in the next generation of colliders \cite{strassler}.

\section{ MeV-scale dark matter and mediators}

The choice of a relatively light  mediator is perhaps the only viable possibility
for WIMP masses to lie well below the Lee-Weinberg window,
and close to the MeV-scale \cite{BF}. This mass range has some 
interesting consequences, including the speculative possibility of linking 
the unexpectedly strong and uniform emission of 511 keV photons from 
galactic center \cite{Integral} to positrons created by the annihilation $O($MeV$)$-scale 
dark matter \cite{511}. To date, much of the model-building in this direction has
 concentrated on utilizing an additional U(1) gauge group, under which both 
 the Standard Model and the dark sector are (disproportionately) charged: a small charge 
 for the Standard Model fermions,  and a larger charge for dark matter \cite{BF,Fayet}. 

The most natural anomaly-free quantum number to gauge is $B-L$ \cite{B-L}, in which case the coupling 
of the additional $U(1)$ gauge bosons to charged fermions is necessarily vector-like. The absence of an 
axial vector current allows this scenario to escape strong constraints from atomic parity violation, and from
flavor-changing decays of $K$ and $B$ mesons \cite{Fayet}. However, there is a price to pay as the coupling to neutrinos
creates a problem with the energetics of supernovae. During the explosion, the 
MeV-scale WIMP is thermalized and coupled to neutrinos too strongly, supressing 
the energy of the emitted neutrinos and making the observed SN1987A signal highly unlikely \cite{SN}. There are
two ways to escape this problem: taking the WIMP mass in excess of 10 MeV, 
or forbidding couplings to neutrinos \cite{SN}. Unfortunately, the first option doesn't work, because 
the shape of the 511 keV line \cite{shape}, along with the $\gamma$-ray spectrum in the MeV region \cite{BY},
do not allow the mass of annihilating particles to be in excess of 3-5 MeV. The second option, i.e. no coupling to 
neutrinos, requires abandoning the initial assumption of gauging $B-L$. 

In this context, it is easy to see that the secluded models of sections 2.1 and 2.2 with vector and scalar mediators
can solve certain model building problems for MeV-scale dark matter. Indeed, neither of these mediation mechanisms
lead to any significant coupling to neutrinos. In particular, for the U(1)$'$ mediator  the $\psi-\nu$ scattering amplitude 
is necessarily suppressed by the $Z$-boson mass, and since the mixing parameter $\kappa$ is smaller than 
$1$, the $\psi-\nu$ scattering cross-section will be below the typical weak-scale value.  This motivates closer inspection
of these two secluded models in the MeV mass range, in order to determine if they do indeed represent viable
MeV-scale WIMP scenarios. In what follows we will make use of two relations that generalize 
results of \cite{Fayet,Ascasibar}, namely for the abundance.
\be
\label{cosmic}
\fr{\Omega_{X}}{\Omega_{DM}} \simeq (2-4)\fr{10^{-36}{\rm cm}^2}{\langle \sigma v\rangle_c },
\ee
and the 511~keV flux, 
\be
\fr{\Phi_{511,X}}{\Phi_{511,{\rm total}}} \sim \fr{ N_{e^+}\langle \sigma v\rangle_g}{10^{-40}{\rm cm}^2}\times 
\left(\fr{1~ {\rm MeV}}{m_X}\right)^2\times
\left(\fr{\Omega_X}{\Omega_{DM}}\right)^2,
\label{galactic}
\ee
where $m_X$ is the mass of the dark matter candidate  $X$ (with 0.5 MeV$ < m_X \la $~3-5 MeV), 
$\Omega_{X}/\Omega_{DM} \leq 1$ is the contribution of $X$ to the total dark matter energy density,
and the ratio $\Phi_{511,X}/\Phi_{511,{\rm total}}$ is the contribution of $X$-annihilation to the total 511 keV $\gamma$-ray flux 
from the galactic center. In (\ref{galactic}),  $N_{e^+}$ is the positron multiplicity per annihilation event. Since the neutrino 
couplings are negligbly small, and direct annihilation to photons is also suppressed \cite{PR2007}, in these models 
$e^+e^-$ {\em is} the dominant  annihilation mode, and $N_{e^+} =1$ for $m_{\rm WIMP} < m_{\rm mediator}$, and $N_{e^+}=2$ otherwise. 
The subscript on the annihilation cross section, $\langle \sigma v \rangle_c$ or  $\langle \sigma v \rangle_g$, denotes the type 
of averaging: $\langle \cdots \rangle_c$ implies thermal averaging for cosmological freezout; while $\langle \cdots \rangle_g$ refers to the average over 
the galactic velocity distribution for the $X$-particles.  We should caution the reader that the relation (\ref{galactic}) 
is only valid to within one to two orders of magnitude, due to the uncertainty in the dark matter number density 
in the galactic core.

\begin{itemize}

\item[(a)] {\bf U(1)$'$-mediator, \boldmath{$m_X > m_V$}}: Irrespective of whether the dark matter is 
a fermion or a scalar, annihilation to two $V$ bosons proceeds in the $s$-channel, and 
$\langle \sigma v \rangle_c\simeq \langle \sigma v \rangle_g$. This immediately rules out MeV-scale 
$X$ particles as the dominant component of dark matter, since $\Omega_{X}/\Omega_{DM} \simeq 1$ 
will ensure that the 511~keV $\gamma$ ray flux is significantly overproduced, $\Phi_{511,X}/\Phi_{511,{\rm total}}\sim 10^4$.
Alternatively, one can tune the coupling $\alpha'$ to be much smaller than the SM gauge couplings, $\alpha'\sim 2\times10^{-3}\times \alpha$, and 
satisfy the 511 keV flux constraint with $\Omega_{X}/\Omega_{DM} \sim 2\times 10^{-5}$. This scenario is reminiscent of 
the decaying sterile neutrino \cite{PP}, where a very small contribution of sterile neutrinos to the 
total dark matter budget can provide the requisite flux. The dominant component of dark matter should 
of course come from other sources, which renders this model incomplete. 

\item[(b)] {\bf U(1)$'$-mediator, \boldmath{$m_X < m_V$}}: In this regime, the choice of $X$ as a scalar charged under 
$U(1)'$ is preferred  \cite{BF}, because of the $p$-wave annihilation that leads to 
$\langle \sigma v \rangle_g\sim 10^{-5}\times  \langle \sigma v \rangle_c$. It then appears entirely possible to satisfy 
both (\ref{cosmic}) and (\ref{galactic}) with $m_X\sim O($MeV$)$, $\Omega_{X}/\Omega_{DM} \simeq 1$, and a 
mixing parameter,
\be
\beta \sim {\rm few} \times 10^{-6}\times \left(\frac{m_V}{10~{\rm MeV}}\right)^4.
\ee
This value for the mixing parameter would be natural for example if induced radiatively at 
a high scale by a state charged both under the SM U(1) and the extra U(1)$'$. As mentioned before, 
this model does not pose any problems with respect to the supernova signal, and does not presuppose 
any hierarchy of gauge couplings as $\alpha'$ can be taken of order $\alpha$. Therefore, this model appears 
the most natural candidate for MeV-scale secluded dark matter, having the chance to explain the 
511~keV line from the galactic center.

\item[(c)] {\bf \boldmath{$\phi$}-mediator, \boldmath{$m_X > m_\phi$}}: In this scenario,
it is advantageous to have a fermionic dark matter candidate $\psi$
with scalar (rather than pseudoscalar) couplings to $\phi$. The annihilation $\psi\psi\to \phi \phi$ proceeeds in the $p$-wave 
and can always be tuned to the required level with a typical choice $\lambda_\psi \sim 10^{-6}$. Since $m_\psi \sim $ few MeV, this 
value of the Yukawa coupling is natural. The subsequent decay of $\phi$ due to mixing with the Higgs is highly suppressed by the 
electron Yukawa coupling,
\be
\Gamma_\phi \sim  \left(\frac{\lambda_1v^2}{m_h^2}\right)^2\times \left(\frac{m_e}{v_{EW}}\right)^2 \times \frac{m_\phi}{8\pi} \ga {\rm sec}^{-1} 
\quad \Longrightarrow \quad  \left(\frac{\lambda_1v^2}{m_h^2}\right)^2 \ga 10^{-8}.
\label{lowerbound}
\ee
The naturalness requirement for the $\ph$-mass would impose a significant constraint here. If we
consider the contribution from Higgs mixing in (\ref{phi-h}), $\lambda_1 v/ m_h \la m_\phi/v$, this
clearly favors a long $\ph$-lifetime ($\sim 1$ sec) and a small mixing parameter.  Even then, 
one must ensure that the ``missing energy" decay $K^+\to \pi^+ +\phi$ is 
within the allowed range. At the quark level, the amplitude for the 
process is given by a Higgs penguin 
(see, {\em e.g.} \cite{PBK}): 
\be
{\cal L}_{eff} = \left(\fr{\lambda_1v^2}{m_h^2}\right) ~\fr{3g_W^2m_sm_t^2V_{td}V^*_{ts}}{64 \pi^2 m_W^2v}~\bar d_Ls_R\phi 
+(h.c.),
\label{Leff}
\ee
leading to the (non-SM) missing energy decay,
\be
\Gamma_{K\to \pi + \ph-{\rm mediator}} \simeq \left(\fr{\lambda_1v^2}{m_h^2}\right)^2
\left(\fr{3m_t^2V_{td}V^*_{ts}}{16 \pi^2 v^2}\right)^2\fr{m_K^3}{64 \pi v^2}.
\ee
Requiring that this width not exceed the observed missing energy decay branching ratio
Br~$=1.5^{+1.3}_{-0.9}\times 10^{-10}$ \cite{Kpinunu} associated with the SM process $K^+\to \pi\nu\bar\nu$, 
results in the following constraint on $\phi-h$ mixing:
\be
\left(\fr{\lambda_1v^2}{m_h^2}\right)^2 < 2\times 10^{-7}.
\label{last_formula}
\ee
This cuts out a significant part of the parameter space, but together with (\ref{lowerbound}) still leaves a 
relatively narrow interval for the mixing parameter, $10^{-7}-10^{-8}$, where the model survives all 
constraints (although not without a modest amount of fine-tuning of the mediator mass) and thus can be the dominant dark matter
component while still accommodating the positron signal through a combination of annihilation and decay. 

The constraints remain essentially the same for a pseudoscalar coupling of $\phi$ to the fermion $\psi$, if the Higgs sector in SM is assumed to be minimal, in 
which case the mixing constant $\lambda_1$ is CP-violating. The additional processes: $s$-wave annihilation $\psi \psi \to e^+ e^-$ through a virtual $\phi$, and also 
 $\psi \psi \to \phi \phi \phi$ if kinematically allowed, are too weak in comparison with the $p$-wave annihilation $\psi \psi \to \phi \phi$ to affect the constraints discussed above. 
In principle, with an extended Higgs sector, $\phi$ could also mix in a CP-conserving way with the physical CP-odd Higgs scalar(s) and in addition could have 
an  enhanced coupling to electrons, relative to the top quark, thus significantly relaxing the constraints on the parameter space in comparison 
with Eqs.~(\ref{lowerbound}) and (\ref{last_formula}).   

\item[(d)] {\bf \boldmath{$\phi$}-mediator, \boldmath{$m_X < m_\phi$}}: In this case the annihilation cross section is suppressed 
by $(m_e/v)^2\sim 4\times 10^{-12}$ and the relation (\ref{cosmic}) would require a mixing parameter, 
$\beta_h$, of order one. Such a choice would involve a gross violation of  naturalness, and would also lead to an unacceptably large 
missing energy signal in $K$ and $B$ decays \cite{PBK}.
We therefore conclude that this option is not viable. 

\end{itemize}

\section{Concluding Remarks}

Secluded WIMP dark matter appears to be a generic possibility, as rather minimal 
model-building choices lead to viable WIMPs interacting with metastable mediators. 
These mediators could be either elementary or composite, and 
we have constructed explicit models with scalar, vector and fermion mediation 
to the Higgs, hypercharge gauge boson, and light neutrino sectors respectively of the SM. 
In the latter case there is a natural choice for the mediator, namely a
right-handed electroweak-scale neutrino. 

Despite existing as a thermal relic, with weak-scale annihilation, none of the secluded WIMP models constructed here 
lead to appreciable signals in underground detectors or register as a missing energy channel in collider 
experiments. Nonetheless, these models are subject to indirect constraints related to 
$\gamma$-rays caused by WIMP annihilation, e.g in the galactic center. Therefore, one of the 
main conclusions of this paper is the complete complementarity of direct and indirect efforts 
for detecting WIMP dark matter.  

Two of the WIMP models constructed here allow a rescaling down to the MeV mass range,
motivated by the intriguing connection to the galactic 511~keV line. 
The secluded models considered here do not couple dark matter to neutrinos, and therefore avoid problems with the suppression
of SN1987A signal. We have shown that models with an additional U(1)$'$ and kinetic mixing to the hypercharge
gauge boson circumvent some problems of scenarios where SM fields carry an additional gauge quantum number,
and appear to be free from unnaturally small parameters. The main limitations come instead from the 
dark matter energy density and from indirect constraints related to the galactic $\gamma$-ray spectrum 
in the MeV region. 

\noindent{\bf Note added --} As this paper was being finalized, we became aware of a recent preprint \cite{kim} that also deals with
U(1)$'$ models of MeV-scale dark matter with kinetic mixing, and thus has some overlap with the discussion in Sect.~3(a,b).

\subsection*{Acknowledgements}

The work of MP and AR was supported in part
by NSERC, Canada. Research at the Perimeter Institute
is also supported in part by the Government of Canada through NSERC and by the Province
of Ontario through MEDT. The work of MV is supported in part by the DOE grant DE-FG02-94ER-40823.

\end{document}